\documentclass[aps,prb,twocolumn,groupedaddress,showpacs]{revtex4}

\usepackage{graphicx}
\usepackage{dcolumn}
\usepackage{bm}

\begin{document}

\title{Magnetic behavior of curium dioxide with non-magnetic ground state}

\author{Fumiaki Niikura$^1$ and Takashi Hotta$^{1,2}$}

\affiliation{$^1$Department of Physics, Tokyo Metropolitan University,
Hachioji, Tokyo 192-0397, Japan \\
$^2$Advanced Science Research Center,
Japan Atomic Energy Agency, Tokai, Ibaraki 319-1195, Japan}

\date{\today}

\begin{abstract}
In order to understand magnetic behavior observed in CmO$_2$
with non-magnetic ground state,
we numerically evaluate magnetic susceptibility on the basis of
a seven-orbital Anderson model with spin-orbit coupling.
Naively we do not expect magnetic behavior in CmO$_2$,
since Cm is considered to be tetravalent ion with six $5f$ electrons
and the ground state is characterized by $J$=0,
where $J$ is total angular momentum.
However, there exists magnetic excited state and
the excitation energy is smaller than the value of
the Land\'e interval rule
due to the effect of crystalline electric field potential.
Then, we open a way to explain magnetic behavior in CmO$_2$.
\end{abstract}



\maketitle


Curium is one of transuranium elements and
it takes trivalent or tetravalent ion state.
It is easy to understand that local $f$ electron number $n$
is equal to 6 and 7 for Cm$^{4+}$ and Cm$^{3+}$ ions, respectively.
By following the well-known Hund's rules,
we obtain that the ground-state multiplet for Cm$^{3+}$
is characterized by $S$=7/2 and $L$=0, while
for Cm$^{4+}$, the multiplet is expressed by $S$=3 and $L$=3,
where $S$ and $L$ denote spin and angular momenta,
respectively.
When we further include the effect of a spin-orbit interaction,
the ground state for Cm$^{3+}$ with $n$=7 is characterized
by $J$=$S$=7/2, which is the same as that for Gd$^{3+}$ ion.
On the other hand, for the case of Cm$^{4+}$ ion with $n$=6,
the ground state is just the singlet with $J$=0.
Thus, the magnetic behavior should be quite different
between Cm$^{3+}$ and Cm$^{4+}$ ions.

Now we pay our attention to experimental results on CmO$_2$,
\cite{Morss}
which are in a puzzling situation.
When we naively deduce the valence of Cm ion
from our experience in other actinide dioxides,
Cm should be tetravalent, leading to non-magnetic ground state.
However, from the neutron diffraction experiment,
the effective moment $\mu_{\rm eff} \sim 3.4\mu_{\rm B}$
has been observed,
where $\mu_{\rm B}$ is the Bohr magneton.
In accordance with the large value of $\mu_{\rm eff}$,
it has been also observed that the magnetic susceptibility
follows the so-called Curie-Weiss behavior.
At the first glance,
these experimental results seem to contradict
non-magnetic ground state of Cm$^{4+}$ ion.

A possibility to resolve the contradiction is to consider
the effect of Cm$^{3+}$ magnetic impurity.
However, the amount of Cm$^{3+}$ deduced from the quantity
of oxygen is apparently smaller than the value to explain
the magnetic moment.
Furthermore, there is no evidence in the
neutron diffraction for the superlattice peak
corresponding to a long-range rearrangement of
oxygen sublattice to accommodate a mixture of
Cm$^{3+}$ and Cm$^{4+}$ ions.
From a theoretical viewpoint,
it is an interesting issue to consider a mechanism to
obtain magnetic behavior from the non-magnetic ground state.

One way to understand the problem
is to consider seriously the magnetic excited state of Cm$^{4+}$ ion,
as already mentioned in the experimental paper.\cite{Morss}
Namely, when the magnetic excited state exists just above the
non-magnetic ground state with a small excitation energy,
it may be possible to understand magnetic behavior
even for the non-magnetic ground state.
This scenario seems to be quite simple,
but we believe that it is worth to
examine faithfully the scenario from a theoretical viewpoint.

In this paper, first we analyze the local $f$-electron state
of Cm$^{4+}$ ion on the basis of the Hamiltonian
including Coulomb interactions, spin-orbit coupling $\lambda$,
and crystalline electric field (CEF) potential.
If we ignore the CEF potential,
the excitation energy $\Delta E$ is given by $\Delta E$=$\lambda/6$
in an $LS$ coupling region,
while it is expressed by $\Delta E$=$7\lambda/2$+$\eta$
in a $j$-$j$ coupling region, where $\eta$ is the energy shift
depending on the CEF potential.
When we further include the CEF potential,
we find the relation of $\Delta E$$\propto$$\lambda^4$
in the $LS$ coupling region,
which is smaller than the value in the Land\'e interval rule.
Then, we evaluate the magnetic susceptibility on the basis of
a seven-orbital Anderson model by exploiting
a numerical renormalization group technique.
It is found that the magnetic behavior actually appears
for a realistic value of $\lambda$ even if we assume Cm$^{4+}$ state.


Let us first discuss the local $f$-electron state.
The model is given by \cite{unit}
\begin{eqnarray}
  && H_{\rm f} \!=\!
  \sum_{m,m'}\sum_{\sigma,\sigma'}
  (B_{m,m'}\delta_{\sigma,\sigma'}
  +\lambda \zeta_{m,\sigma,m',\sigma'})
  f_{m\sigma}^{\dag}f_{m'\sigma'} \nonumber \\
  && +\sum_{m_1 \sim m_4}\sum_{\sigma,\sigma'}
  I_{m_1m_2,m_3m_4}
  f_{m_1\sigma}^{\dag}f_{m_2\sigma'}^{\dag}
  f_{m_3\sigma'}f_{m_4\sigma},
\end{eqnarray}
where $B_{m,m'}$ is the CEF potential,
$\sigma$=$+1$ ($-1$) for up (down) spin,
$f_{m\sigma}$ is the annihilation operator for $f$ electron with
spin $\sigma$ and $z$-component $m$ of angular momentum $\ell$=3,
$\delta_{\sigma\sigma'}$ is the Kronecker's delta,
$\lambda$ is the spin-orbit coupling,
$\zeta_{m,\pm 1,m,\pm 1}=\pm m/2$,
$\zeta_{m \pm 1,\mp 1,m, \pm 1}=\sqrt{12-m(m \pm 1)}/2$,
and zero for the other cases.
The Coulomb integral $I$ is expressed by the combination of
Slater-Condon parameters such as $F^0$, $F^2$, $F^4$, and $F^6$.
\cite{Slater}


Here we briefly explain the parameters of the local Hamiltonian
$H_{\rm f}$.
Concerning Slater-Condon parameters,
first we set $F^0$=10 eV by hand,
since we are not interested in the determination of
the absolute value of the ground state energy.
Others are determined so as to reproduce excitation spectra of
U$^{4+}$ ion with two $5f$ electrons.\cite{Eliav}
Here we show only the results:
$F^2$=6.36 eV, $F^4$=5.63 eV, and $F^6$=4.13 eV.
As for spin-orbit coupling $\lambda$,
it may be possible to use an experimental value,
but in this paper,
we change the value of $\lambda$ as a parameter to control
the situation from the $LS$ coupling to $j$-$j$ coupling schemes.
Since the fluorite structure belongs to $O_{\rm h}$ point group,
$B_{m,m'}$ is given by using a couple of CEF parameters,
$B_4^0$ and $B_6^0$, which are traditionally expressed as
$B_4^0$=$Wx/15$ and $B_6^0$=$W(1-|x|)/180$,
respectively.\cite{LLW,Hutchings}
Here $W$ determines the energy scale of the CEF potential,
while $x$ specifies the CEF scheme.

Now we summarize the experimental facts on the CEF energy levels
of actinide dioxides:
For UO$_2$, the ground state is $\Gamma_5^+$ triplet and
the first excited state is $\Gamma_3^+$ doublet with the
excitation energy 150 meV.\cite{Amoretti}
For NpO$_2$, the ground and first excited states are,
respectively, $\Gamma_8^{-(2)}$ and $\Gamma_8^{-(1)}$
quartets with the excitation energy 55 meV.\cite{Fournier}
For PuO$_2$, the ground state is $\Gamma_1^+$ singlet,
while the first excited state is $\Gamma_4^+$ triplet
with the excitation energy 123 meV.\cite{Kern1,Kern2}
In order to reproduce totally the above CEF level schemes,
we have chosen $W$=$-10.5$ meV and $x$=0.62
in the discussion of the ground state of AmO$_2$.\cite{Hotta1}

\begin{figure}[t]
\includegraphics[width=8.5cm]{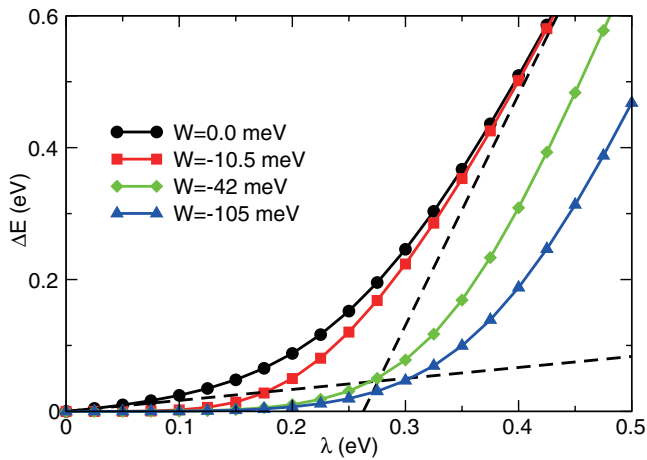}
\caption{
Excitation energy $\Delta E$ vs. spin-orbit coupling $\lambda$
for $n$=6.
Thick solid curves denote the calculated results with different values
of the CEF potential $W$.
Broken lines indicate the two limiting situations as
$\Delta E$=$\lambda/6$ for small $\lambda$ and
$\Delta E$=$7\lambda/2+\eta$ for large $\lambda$,
where $\eta$ is the energy shift depending on the CEF potential.
}
\end{figure}


In Fig.~1, we show the excitation energy $\Delta E$ vs. $\lambda$
for various values of $W$.
Note that at $\lambda$=0, the ground state is highly degenerate.
Here we set $\Delta E$=0 at $\lambda$=0 in order to visualize
how the degeneracy is lifted by the inclusion of $\lambda$.
First we consider the case of $W$=0, in which the ground state is
characterized by $J$=0, while the first excited state is given by $J$=1.
In this case, we obtain $\Delta E$=$\lambda/6$ in the $LS$
coupling region from the so-called Land\'e interval rule.
On the other hand, in the $j$-$j$ coupling region,
the ground state is expressed by the closed shell of $j$=5/2 sextet
and the exited state is given by the one-electron excitation
from $j$=5/2 sextet to $j$=7/2 octet, leading to the $J$=1 excited state.
Thus, we obtain $\Delta E$=$7\lambda/2$+$\eta$ in the $j$-$j$ coupling scheme,
where $\eta$ denotes the energy shift depending on the CEF potentials.

Now we include the effect of the CEF potential.
As shown in Fig.~1, the excitation energy becomes small
when we set $W$=$-10.5$ meV, which has been determined so as
to reproduce the CEF energy levels of other actinide dioxides.
Both the ground state and the first excited state energies
are decreased by the CEF potential, but the reduction of the
first excited state energy is large in comparison with
that of the ground state energy.
From the analysis of our numerical results, we find the relation
of $\Delta E \propto \lambda^4$, suggesting that $\Delta E$ is
suppressed for small $\lambda$ in comparison with the relation of
$\Delta E$=$\lambda/6$ for $W$=0.
When we further increase the magnitude of $W$, the excitation
energy is getting smaller and smaller, as observed in Fig.~1.
Thus, it is concluded that the excitation energy of
the $f^6$ electron system is smaller than we have naively expected
due to the combined effect of Coulomb interactions,
spin-orbit interaction, and the CEF potential.


Let us now include the hybridization between localized and
conduction electrons.
The Hamiltonian is the seven-orbital Anderson model,
\cite{Hotta2} given by
\begin{equation}
  H \!=\! \sum_{{\bf k},\sigma}
  \varepsilon_{{\bf k}} c_{{\bf k}\sigma}^{\dag} c_{{\bf k}\sigma}
  +\sum_{{\bf k},\sigma,m}
  (V_{m} c_{{\bf k}\sigma}^{\dag}f_{m\sigma}+{\rm h.c.})
  +H_{\rm f},
\end{equation}
where $\varepsilon_{\bf k}$ denotes conduction electron dispersion,
$c_{{\bf k}\sigma}$ indicates the annihilation operator for conduction
electron with momentum ${\bf k}$ and spin $\sigma$,
and $V_{m}$ is the hybridization between conduction and $f$ electrons.

Note that we consider only $a_{\rm u}$ single conduction band
with xyz symmetry composed of oxygen $2p$ electrons.
Since oxygen ions surrounding actinide ions are located
in the [1, 1, 1] direction, there should exist a conduction band
composed of $2p$ electrons with xyz symmetry.
This picture seems to be consistent with band-structure
calculation,\cite{Maehira}
but we assume the ignorance of $t_{\rm 1u}$ and $t_{\rm 2u}$ bands.
Here we note that the hybridization occurs between the states
with the same symmetry of local $f$-electron state.
Since the $a_{\rm u}$ conduction band has xyz symmetry,
we set $V_2$=$-V_{-2}$=$V$ and zero for other $m$,
where $V$ is fixed as $V$=0.05 eV.
In order to adjust the local $f$-electron number $n$,
we appropriately change the chemical potential in the
actual calculation,
although we do not explicitly show such a term.
A half of the bandwidth of $a_{\rm u}$ conduction band
is set as 1 eV, which is the energy unit in the following
calculations.


In this paper, we analyze the seven-orbital Anderson model
by using the numerical
renormalization group (NRG) technique.\cite{Wilson,NRG}
In the NRG calculations, in order to consider efficiently the
conduction electrons near the Fermi energy, the momentum space
is logarithmically discretized and the conduction electron states
are characterized by ``shell'' labeled by $N$.
The shell of $N$=0 denotes an impurity site including $f$ electrons.
The Hamiltonian is transformed into the recursion form as
\begin{eqnarray}
  H_{N+1} = \sqrt{\Lambda}H_N+t_N \sum_\sigma
  (c_{N\sigma}^{\dag}c_{N+1\sigma}+c_{N+1\sigma}^{\dag}c_{N\sigma}),
\end{eqnarray}
where $\Lambda$ is a parameter for logarithmic discretization,
$c_{N\sigma}$ denotes the annihilation operator of conduction electron
in the $N$-shell, and $t_N$ is ``hopping'' of electron between $N$-
and $(N+1)$-shells, given by
\begin{eqnarray}
  t_N=\frac{(1+\Lambda^{-1})(1-\Lambda^{-N-1})}
  {2\sqrt{(1-\Lambda^{-2N-1})(1-\Lambda^{-2N-3})}}.
\end{eqnarray}
The initial term $H_0$ is given by
\begin{eqnarray}
  H_0=\Lambda^{-1/2}[H_{\rm f} + \sum_{m\sigma}
  V_{m}(c_{0\sigma}^{\dag}f_{m\sigma}+f_{m\sigma}^{\dag}c_{0\sigma})].
\end{eqnarray}
In this paper, $\Lambda$ is set as $5$ and we keep $4500$
low-energy states for each renormalization step.

\begin{figure}[t]
\includegraphics[width=8.5cm]{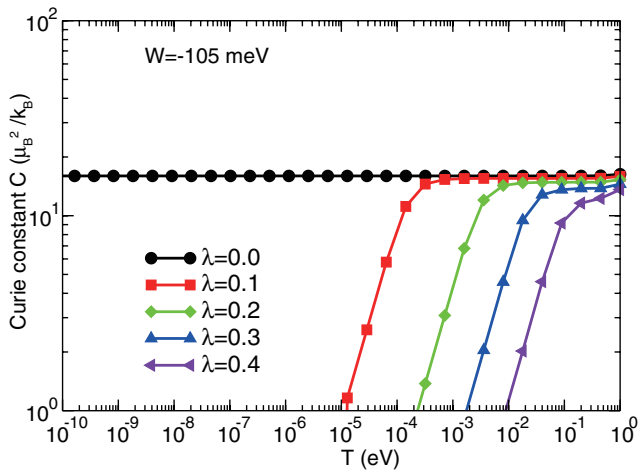}
\caption{Curie constant $C$ vs. temperature $T$
for various values of $\lambda$.
}
\end{figure}

In order to see magnetic properties, we evaluate
the magnetic susceptibility of $f$-electron, defined by
\begin{eqnarray}
   \chi \!=\! \frac{1}{T} \lim_{N \rightarrow \infty}
   \Biggl[ \frac{{\rm Tr} M_{z,N}^2 e^{-H_N/T}}{{\rm Tr} e^{-H_N/T}}
   - \frac{{\rm Tr} S_{z,N}^2 e^{-H^0_N/T}}{{\rm Tr} e^{-H^0_N/T}}
   \Biggr],
\end{eqnarray}
where $T$ denotes a logarithmic temperature given by $T$=$\Lambda^{-(N-1)/2}$
in the NRG calculation, $H_N^0$ is the Hamiltonian without $H_{\rm f}$,
and a magnetic moment along the $z$-axis is defined by
\begin{eqnarray}
   M_{z,N}=\mu_{\rm B}\sum_{m,\sigma}(m+g_s\sigma/2)
   f_{m\sigma}^{\dag}f_{m\sigma}+S_{z,N}.
\end{eqnarray}
Here $\mu_{\rm B}$ is the Bohr magneton, $g_s$ is electron $g$-factor
given by $g_s$=2, and
\begin{eqnarray}
  S_{z,N}=g_s\mu_{\rm B}\sum_{n=0}^{N}\sum_{\sigma}(\sigma/2)
   c_{n\sigma}^{\dag}c_{n\sigma}.
\end{eqnarray}
Note that we consider only the susceptibility for
the $z$-component of the magnetic moment, but
we have performed the calculations for all directions
of the magnetic moment.
Then, we have checked that the susceptibilities for the
magnetic moments along $x$-, $y$-, and $z$-directions
take the same values due to the cubic symmetry.
In the following, we show only the results for the susceptibility
for the magnetic moment along the $z$-axis.

In Fig.~2, we show the Curie constant $C$ as a function of $T$.
In NRG calculations, it is possible to evaluate precisely
the Curie constant $C$ by using the relation $T \chi$.
We note that the finite value of $C$ suggests the emergence of
of magnetic behavior even for the non-magnetic ground state.
Here the CEF potential is chosen as $W$=$-105$ meV in order to
obtain the small excitation energy.
Concerning the large magnitude of $W$,
we will provide a brief comment later.

\begin{figure}[t]
\includegraphics[width=8.5cm]{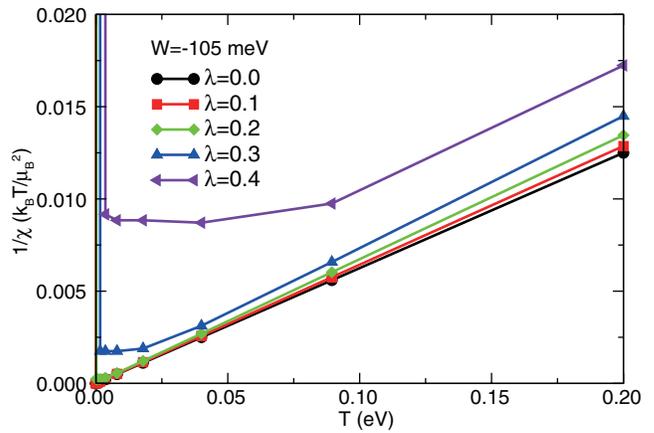}
\caption{Inverse susceptibility $1/\chi$ vs. temperature $T$
for various values of $\lambda$.
The CEF potential is set as $W$=$-105$ meV.
}
\end{figure}

First let us consider the case of $\lambda$=0,
in which we find $C$=$16 \mu_{\rm B}^2/k_{\rm B}$
almost irrespective of a temperature, as observed in Fig.~2.
Note that the residual entropy from the localized magnetic moment
is due to the speciality of the present model
in which we consider only the hybridization
with single $a_{\rm u}$ band.
If we further include $t_{\rm 1u}$ and $t_{\rm 2u}$ bands,
the residual entropy will be released due to the Kondo effect.

Now we consider the meaning of ``16" in the value of $C$ for $\lambda$=0.
In this situation, the ground state is characterized by $S$=3 and $L$=3
due to the Hund's rules,
indicating 49-fold degeneracy in the ground state.
When we include the effect of CEF potential,
we note that the CEF potentials act on the charge,
not on the spin.
Namely, the septet of $L$=3 splits into one single and two triplets
due to the effect of the cubic CEF potentials.
For $W<0$ and $0<x<1$, the lowest energy state is
found to be the singlet.
Thus, the ground state becomes septet characterized by $S$=3.
As is well known, the Curie constant of the susceptibility of
the magnetic moment $M$=$L+g_s S$=$g_J J$ is given by
$C$=$g_J^2 \mu_{\rm B}^2 J(J+1)/(3k_{\rm B})$,
where $g_J$ is the Land\'e's $g$-factor.
In the present case, $J$ is composed only of electron spins
and thus, we set $J$=$S$=3 and $g_J$=$g_s$=2,
leading to $C$=$16 \mu_{\rm B}^2/(3k_{\rm B})$.
Such a large value of $C$ at $\lambda$=0 is related to
the appearance of magnetism for $\lambda>0$.

When we increase the value of $\lambda$,
$C$ should be decreased,
since the ground state becomes the singlet
characterized by $J$=0.
However, as shown in Fig.~2, the Curie constant does $not$
vanish immediately, when we increase the value of $\lambda$.
In fact, as observed in the curve for $\lambda$=0.1,
$C$ is still constant in the temperature region of $T \agt 10^{-4}$
due to the smallness of the excitation energy.
When $\lambda$ is further increased, the temperature
dependence of the Curie constant becomes significant.
For $\lambda$=0.4, $C$ becomes small drastically
at a temperature as large as $T \sim 10^{-2}$.

In order to compare the numerical results with the experimental ones,
we plot the inverse susceptibility $1/\chi$ in Fig.~3.
Here we use a large value for CEF potential as $W$=$-105$meV,
in order to reproduce the small excitation energy.
We note that $1/\chi$ is shown in the linear scale of the temperature,
even though the low-temperature precise behavior
in the logarithmic-scale calculation cannot be observed
in the linear scale for the temperature.
Note also that except for $\lambda$=0, $1/\chi$ diverges near $T \sim 0$,
since the ground state is singlet and the Curie constant eventually
goes to zero at low temperatures.
In the expression of $1/\chi$ in the linear scale, 
even for $\lambda$=0.3,
we find the region of Curie-Weiss-like behavior in a temperature region
of $T \agt 0.01$.
For $\lambda$=0.4, it is difficult to observe the Curie-Weiss-like
behavior in the temperature range of $T < 0.1$.
From these results, we conclude that the magnetic behavior of CmO$_2$
can be explained even for the non-magnetic ground state of Cm$^{4+}$
if we appropriately choose the parameters in the model.


In the present scenario, the magnetic behavior should disappear
at low enough temperatures.
However, it will be difficult to perform low-temperature experiments
for CmO$_2$.
An executable test is to measure the temperature dependence of
the magnetic moment in the neutron diffraction experiment.
If the present scenario is correct, it is possible to detect the
reduction of the moment due to the decrease of temperature.

Finally, we briefly comment on the value of $W$.
In the present calculation for $\chi$, the value of $W$ may be
too large in comparison with that for other actinide dioxides.
\cite{Hotta1}
Namely, our result indicates that such a large CEF potential is
required to explain the magnetic behavior in CmO$_2$ on the basis
of the model calculations.
A possible detection of the large CEF effect in CmO$_2$
will be another test for the present scenario.


In summary, we have analyzed the magnetic susceptibility of CmO$_2$
on the basis of the seven-orbital Anderson model by using the
NRG technique.
We have actually found that the magnetic behavior appears
even if the ground state of Cm ion is non-magnetic,
since the excitation energy between ground and magnetic excited states
becomes small due to the combined effect
of Coulomb interactions, spin-orbit coupling, and CEF potential.
We believe that the scenario works to understand the magnetism in CmO$_2$.


The authors thank S. Kambe and Y. Tokunaga for discussions on
actinide dioxides.
This work has been supported by a Grant-in-Aid for
for Scientific Research on Innovative Areas ``Heavy Electrons''
(No. 20102008) of The Ministry of Education, Culture, Sports,
Science, and Technology, Japan.
The computation in this work has been done using the facilities
of the Supercomputer Center of Institute for Solid State Physics,
University of Tokyo.



\begin{thebibliography}{99}
\bibitem{unit}
In this paper, we use such units as $\hbar$=$k_{\rm B}$=1.
\end{thebibliography}
\end{document}